\documentclass[ amsmath, amssymb,10pt,sort&compress ]{article}
\usepackage{amsmath,amssymb,graphicx}
\usepackage{xcolor}
\usepackage{hyperref}
\usepackage{cleveref}
\usepackage{filecontents}
\usepackage{authblk}
\usepackage{graphicx}

\title{Fermionic Tachyons as a Source of Dark Energy}

\author[1]{Salomeh Khoeini-Moghaddam\thanks{skhoeini@khu.ac.ir}}
\author[1]{Farzan Momeni\thanks{momeni@khu.ac.ir }}
\author[1]{Fatemeh Yousefabadi\thanks{fatemeh.yousefabadi@gmail.com}}
\affil[1]{Department of Astronomy and High Energy Physics, Faculty of Physics, Kharazmi University, Mofateh Ave., Tehran 15719-14911, Iran}

\begin{document}

\maketitle

\begin{abstract}
We investigate a model for the universe based on the self-interacting fermionic tachyon field. We have shown that by devising a self-interaction potential of a proper form, the fermionic tachyon field is capable of producing an accelerating expansion.  At late times, this potential tends to a constant value; this behavior is consistent with the cosmological constant. In this way, the introduced fermionic tachyon field can be interpreted as the source of dark energy.


\end{abstract}

\section{Introduction}
Modern cosmology deals with two periods of accelerating expansion. The first is an assumed inflationary era elaborated to explain some of the main problems of standard cosmology, mainly as a means to account for perturbation seeds, and of course, to resolve flatness and horizon problems. The second period is the present era: our universe expands with a positive acceleration rate~\cite{1998AJ116.1009R,1999ApJ517565P}.
 In this work, our concern will be with the latter period.  Several models have been devised to explain the accelerating universe. The simplest model is the $\Lambda$CDM model, in which a constant is added to the Einstein field equations to deputize the accelerating expansion of the universe. This constant is called the cosmological constant. It also deputizes our ignorance about the physics of dark energy.
 The $\Lambda$CDM model, while being in  good agreement with observational data~\cite{ade2016planck,2018planck}, however, lacks the phenomenology. By far, the physical agent, namely the dark energy, that accelerates the expansion of our universe remains unknown.

  According to a large amount of observational data~\cite{huterer2017dark, lonappan2018bayesian}, dark energy contributes about 68 percent of the total energy in our present universe.
 There are many different approaches towards the explanation of dark energy ~\cite{de2010f,amendola2010dark,carroll2001cosmological, copeland2006,lee2008palatini,peebles2003cosmological, padmanabhan2003cosmological, Tsujikawa2013, Bamba2012,NOJIRI201159}. One of these approaches is to investigate some (unusual) matter which, whatsoever, could be responsible for the accelerating expansion of the universe.
It is also shown that a fermionic source can lead to an accelerating expansion ~\cite{obukhov1993spin,saha1997interacting,saha2001spinor,saha2004bianchi, saha2006nonlinear,armendariz2003spinors,de2008noether,samojeden2010fermions,grams2014fermion,ribas2005fermions,ribas2007cosmological,ribas2011fermionic,
ribas2012fermions,ribas2015isotropization,ribas2016cosmological,ribas2017fermionic,Jamil2012,Kucukakca2014}.

 Several models for fermionic fields are investigated. It is shown that in these models, a fermionic field takes a tremendous role in both the early universe inflation and the late-time universe dark energy. As inspired by the string theory,  a tachyon field also produces inflation in the very early universe ~\cite{gibbons2002cosmological,jentschura2012pseudo,saha2016nonlinear,salesi1997slower,bandukwala1974theory}.
  Also, it is shown that a scalar tachyon field can be interpreted as dark energy ~\cite{martins2016cosmological,bagla2003cosmology}.

In general, tachyons can have any spin\cite{schwartz2021}; in particular, they can have
spin 1/2[\cite{schwartz2016}. In this work, we investigate whether a fermionic tachyon can also take the role of dark energy. The idea of the existence of the fermionic tachyon field is inspired by some evidence that neutrinos, as we expect from fermionic tachyons ~\cite{chodos1985neutrino,schwarts2017}, may have an effective negative mass ~\cite{EHRLICH201511,nanni2020} and references therein.
 Of course, the fermionic tachyon field which takes the role of dark energy in our model is not necessarily a neutrino field.
We consider the universe to include two fields; a matter field that naturally tends to decelerate the universe's expansion and a fermionic tachyon field that overwhelmingly accelerates its expansion.
We assume that there is no local interaction between these two components; however, they feel each other only vie their overall gravity. We also take account of the irreversible processes between matter and gravity to thermalize the expanding universe.
A relativistic quantum fermion is describable by the Dirac equation.
 So, first of all, we should think of devising a modified form of this equation to describe fermionic tachyons. Insequent, we use tetrad formalism to establish a link between the modified Dirac equation and general relativity. Then, we solve the resultant Einstein's equations of motion with a self-interaction potential in a flat FRW background.
 We choose a specific form for the potential and solve these equations numerically.  The signature (+,-,-,-) and natural units, i.e. $8\pi G=c=\hbar=k=1$ are used.

 The tetrad and Dirac formalism for tachyons in curved space-time is reviewed in Sec.\ref{Dirac eq}. The field equations for a spatially flat isotropic and homogeneous universe in the presence of a fermionic tachyon field are derived and solved numerically in Sec.\ref{Dirac FRW}. We conclude and discuss our results in Sec.\ref{conclusion}.


\section{Review of Tetrad Formalism and Fermionic Tachyon Lagrangian}\label{Dirac eq}
 In this section, we review the techniques used to include fermions in general relativity~\cite{wald2010general, ryder1996quantum,birrell1984quantum,weinberg2014gravitation} and then generalize the result into tachyons. Of course, the spinor representation does not appear firsthand in the gauge group of general relativity.
However, the tetrad formalism is capable of providing a means to generalize the Dirac equation, which was originally written for a Minkowski space-time, to curved space-time. In this approach, the normal coordinate at each space-time point is constructed. In this coordinate, the metric is  $\eta_{ab}$. The curved space-time metric, $g_{\mu\nu}$ can be obtained via the  relation,
$g_{\mu\nu}=e^a_\mu e^b_\nu\eta_{ab}$.

Here $e^{\mu}_{}$ is tetrad or vierbein which, as imposed by general covariance principle, relates quantities in the local inertial frame to their counterparts in curved space-time frame. The Latin and Greek indices refer to Minkowski and curved space-time, respectively.
The Dirac Lagrangian in Minkowski space-time is
\begin{align}
L_D=\frac{i}{2}[\bar{\psi}\gamma^a\partial_a\psi-\left(\partial_a\bar{\psi}\right)\gamma^a\psi]-m\bar{\psi}\psi-V,
\end{align}
 where  $\bar\psi=\psi^\dagger\gamma^0$ is the Dirac adjoint spinor field, $m$ denotes the fermionic mass, and  $V$ represents the potential density of self-interaction between fermions which is a function of $\psi$ and $\bar\psi$.

As imposed by this formalism, we should replace the Dirac--Pauli matrices $\gamma^{a}$ with their counterparts in curved space time~\cite{birrell1984quantum}, as

$\Gamma^{\mu}=e^{\mu}_{a}\gamma^{a}$.

 These generalized Dirac--Pauli matrices satisfy $\{\Gamma^\mu,\Gamma^\nu\}=2g^{\mu\nu}$. We also need to substitute the generalized covariant derivatives for their ordinary derivatives, i.e.
\begin{align}
\partial_{\mu}\psi\longrightarrow D_{\mu}\psi=\partial_{\mu}\psi-\Omega_{\mu}\psi,\quad
\partial_{\mu}\bar{\psi}\longrightarrow D_{\mu}\bar{\psi}=\partial_{\mu}\bar{\psi}+\bar{\psi}\Omega_{\mu}
\label{covdev}
\end{align}
where  $\Omega_{\mu}$,s are spin connections defined as
$\Omega_{\mu}=-\dfrac{1}{4}g_{\rho\sigma}\left[ \Gamma^{\rho}_{\mu\delta}-e_{b}^{\rho}\left( \partial_{\mu}e^{b}_{\delta}\right) \right]  \Gamma^{\sigma}\Gamma^{\delta},$

with $\Gamma^{\rho}_{\mu\delta}$ denoting the Christoffel symbols.
 By replacing the covariant quantities in Dirac Lagrangian, we arrive at the "generally covariant Diac Lagrangian":
 \begin{align}
\mathcal{L}_{D}=\frac{i}{2}[\bar{\psi}\Gamma^\mu D_\mu\psi-\left(D_\mu\bar{\psi}\right)\Gamma^\mu\psi]-m\bar{\psi}\psi-V.
\label{Dirac Lagrangian}
\end{align}

Now we require to write a Lagrangian for a fermionic (pseudo) tachyon in a locally inertial frame and then transform the results into the general frame. According to Weyl constraint for massless spinors, we have
\begin{align}
\gamma_5\psi=-\psi.
\end{align}
The above equation admits to writing the kinetic part as two forms that are indistinguishable when the mass and potential terms are absent,
\begin{align}
L_{K}=\frac{i}{2}[\bar{\psi} \gamma^{a} \partial_{a}\psi -\left(\partial_a\bar{\psi}\right)\gamma^a \psi]
\end{align}
\begin{align}
\tilde{L}_{K}=\frac{i}{2}[\bar{\psi}\gamma^{5} \gamma^{a} \partial_{a}\psi -\left(\partial_a\bar{\psi}\right)\gamma^{5}\gamma^a \psi].
\end{align}
 In the presence of mass and potential term ( due to some short distance interaction), the first equation belongs to the normal Dirac fermion and the latter one describes the kinetic part of the fermionic tachyon. In the context of field theory, the tachyon is a field with a negative square mass. The addition of the mass and potential terms can form a Lagrangian for tachyon\cite{chodos1985neutrino}. In summary,
the Dirac Lagrangian density for a  (pseudo) fermionic tachyon in Minkowski  space-time can be written as ~\cite{chodos1985neutrino,jentschura2012pseudo,salesi1997slower,jentschura2013generalized}
\begin{align}
\tilde{L}_{D}=\frac{i}{2}[ \bar{\psi}\gamma^{5} \gamma^{a} \partial_{a}\psi -\left(\partial_a\bar{\psi}\right)\gamma^{5}\gamma^a \psi]- m\bar{\psi}\psi-V.
\label{tachyon lag}
\end{align}

To generalize this equation to curved space-time, we follow the tetrad formalism and replace the covariant quantities
in tachyon, Lagrangian to arrive at the "covariant tachyon Dirac Lagrangian" as below
\begin{align}
\tilde{\mathcal{L}}_{D}=\frac{i}{2}[\bar{\psi}\Gamma^{5} \Gamma^{\mu} D_{\mu}\psi-\left(D_\mu\bar{\psi}\right)\Gamma^{5}\Gamma^\mu \psi]-m\bar{\psi}\psi -V,
\label{ tachyon covariant Lagrangian}
\end{align}
with $\Gamma^5=-i\sqrt{-g}\Gamma^0\Gamma^1\Gamma^2\Gamma^3$, the Euler-Lagrange equations give the "covariant Dirac equations" for the tachyonic spinor field and its adjoint coupled to gravity as
\begin{align}
\begin{split}
&i\Gamma^5\Gamma^\mu D_\mu\psi- m\psi - \dfrac{dV}{d\bar{\psi}}=0\\
&iD_\mu\bar{\psi}\Gamma^5\Gamma^\mu +m\bar{\psi}+ \dfrac{dV}{d\psi} =0.
\end{split}
\label{eqm1 Tachyon}
\end{align}
Compared to corresponding equations for ordinary fermions, we have an extra $\Gamma^5$ besides $\Gamma^\mu$.

Meanwhile, the total action is
\begin{align}
S(g,\psi,\bar{\psi})=\int d^4x\sqrt{-g}(\tilde{\mathcal{L}}_{D}+\tilde{\mathcal{L}}_m+\tilde{\mathcal{L}}_g),
\end{align}
where $\tilde{\mathcal{L}}_m$ is the matter Lagrangian density, and $\tilde{\mathcal{L}}_g=\frac{R}{2}$ is the gravity part (R is the Ricci scalar).
Variation of the total action concerning metric  $g^{\mu\nu}$ gives Einstein's equations,
\begin{align}
R_{\mu\nu}-\frac{1}{2}g_{\mu\nu}R=-T_{\mu\nu},
\end{align}
in which $T_{\mu\nu}$ is the total energy-momentum $T_{\mu\nu}=(\tilde{T}_D)_{\mu\nu}+(T_m)_{\mu\nu}$. The first term is due to the fermionic tachyon field, and the second term is related to the matter field;
\begin{align}
\begin{split}
 (\tilde{T}_f) _{\mu \nu}=\frac{i}{4}[\bar{\psi}\Gamma^5\Gamma_\mu D_\nu\psi+\bar{\psi}\Gamma^5\Gamma_\nu D_\mu\psi-D_\mu\bar{\psi}\Gamma^5\Gamma_\nu\psi-D_\nu\bar{\psi}\Gamma^5\Gamma_\mu\psi]-g_{\mu\nu}\tilde{\mathcal{L}}_D.
 \end{split}
\label{T_f}
\end{align}

\section{Dirac and Einstein Equations in FRW Background}\label{Dirac FRW}
 To be more explicit, we consider a spatially flat FRW metric that is consistent with observational data. The metric is as
$ds^2=dt^2-a(t)^2(dx^2+dy^2+dz^2)$
where $a(t)$ is the scale factor.
 The energy-momentum tensor for perfect fluid is as follows,
\begin{align}
(T^{\mu}_\nu)=diag(\rho, -p, -p, -p),
\label{energy momentum}
\end{align}
 where $\rho$ and $p$ denote the sum of energy density and the pressure of fermionic tachyon and ordinary matter field, respectively.
 As usual, the Friedman equations in flat FRW metric are
\begin{align}
H^2 =\frac{1}{3} \rho
\label{FRW1}
\end{align}
\begin{align}
\frac{\ddot{a}}{a} =-\frac{1}{2}H^2-\frac{1}{2}p.
\label{FRW2}
\end{align}
 with $H = \dot{a}(t)/{a(t)}$  being the Hubble parameter. For later convenience, we decompose the tachyon and the matter contributions, i.e.  $\rho=\rho_{f}+\rho_{m}$ and $p=p_{f}+p_{m}$.
 The non-vanishing components of the energy-momentum tensor of the fermionic tachyon field  (\ref{T_f}) are as follows,
\begin{align}
\rho_f &=m\bar{\psi} \psi + V\nonumber\\
p_f &=\frac{1}{2}\dfrac{dV}{d\psi} \psi + \frac{1}{2}\bar{\psi}\frac{dV}{d\bar\psi} - V.
\label{energy and pressure}
\end{align}

We use the tetrad formalism with this metric; the tetrad components are as follows,
\begin{align}
e_0 ^\mu = \delta _0 ^\mu, && e_ i ^\mu = \frac{1}{a(t)}\delta _i ^\mu.
\label{FRW tetrad}
\end{align}
Hence the Dirac matrices and the spin connection components, respectively, become
\begin{align}
\Gamma^0 =\gamma^0 ,\quad \Gamma^i = \dfrac{1}{a(t)}\gamma^i ,\quad \Gamma^5 = -i\sqrt{-g}\Gamma^0 \Gamma^1 \Gamma^2 \Gamma^3 =\gamma^5
\label{FRW Drirac}
\end{align}
and
\begin{align}
\Omega_0 =0, \quad \Omega _i= \frac{1}{2} \dot{a}(t)\gamma ^i \gamma ^0.
\label{FRW connection}
\end{align}
With the above relations, the Euler-Lagrange equations for fermionic tachyon (\ref{eqm1 Tachyon}) are simplified as
\begin{align}
\begin{split}
&\gamma^5\dot{\psi} + \dfrac{3}{2} H \gamma^5 \psi - im\gamma^0 \psi - i \gamma^0 \dfrac{dV}{d\bar{\psi}}=0\\
&\dot{\bar{\psi}} \gamma^5 + \dfrac{3}{2} H \bar{\psi}\gamma^5 - im\bar{\psi}\gamma^0 - i  \dfrac{dV}{d\psi} \gamma^0=0.
\end{split}
\label{FRW eqm}
\end{align}
The continuity equation is given by the conservation law of total energy-momentum tensor $T^{\mu\nu}_{;\nu}=0$ as
\begin{align}
\dot{\rho}+3H(\rho + p) = 0,
\label{cons}
\end{align}
but the conservation law for the energy density of the tachyon field can be obtained from (\ref{FRW eqm}) and (\ref{energy and pressure}) as
\begin{align}
\dot{\rho_f}+3H(\rho_f + p_f) = 0.
\label{fermion cons}
\end{align}
That means the conservation equation of the tachyon field is decoupled from the matter field; i.e. the density evolution of the fermionic tachyon field is decoupled from the matter density. From (\ref{cons}) and (\ref{fermion cons}) for the matter field density we have
\begin{align}
\dot\rho_m+3H(\rho _m + p_m) =0.
\label{matter cans}
\end{align}
We consider a barotropic equation of state for the matter field, i.e.  $p_{m}=w_{m}\rho_{m}$ with $0\leq w_{m}\leq1$.

 In order to get cosmological solution we must specify the exact form of the self-interaction  potential "$V$" which is a function of scalar invariant $(\bar{\psi}\psi)^2$ and pseudo-scalar invariant $(\bar{\psi}\gamma^5\psi)^2$. We consider a self-interaction potential of the  below form ~\cite{ribas2005fermions,saha2016nonlinear,grams2014fermion,Inagaki1997,Rantaharju2017,INAGAKI2017297,Kucukakca2015},
\begin{align}
V = \lambda [\beta_1 (\bar{\psi} \psi)^2 + \beta_2 (i\bar{\psi} \gamma^5 \psi)^2]^n,
\label{v}
\end{align}
where $\lambda$ is a non-negative coupling constant that denotes the strength of potential. Moreover $\beta_1$ and $\beta_2$ are constants specifying the role of scalar and pseudo-scalar. We choose $\beta_1=\beta_2=1$  all over this work. The exponent $n$ is another free parameter in this model.

By substituting (\ref{v}) in (\ref{energy and pressure}), we have
\begin{align}
\rho_f =m(\bar{\psi} \psi) + \lambda [\beta_1 (\bar{\psi} \psi)^2 + \beta_2 (i\bar{\psi} \gamma^5 \psi)^2]^n,\\
p_f = (2n-1) \lambda [\beta_1 (\bar{\psi} \psi)^2 + \beta_2 (i\bar{\psi} \gamma^5 \psi)^2]^n.
\label{fermion p}
\end{align}
Since we are interested in the case of fermionic tachyon taking the role of dark energy, $p_f$ must be negative. Hence we would have $n<\frac{1}{2}$. This way, the equation of motion of fermionic tachyon (\ref{FRW eqm}), the Friedman equations (\ref{FRW1} and \ref{FRW2}), the matter conservation equation (\ref{matter cans}),  form a system of coupled ordinary differential equations. In order to obtain the cosmological quantities in terms of red shift, $z=\frac{a_0}{a(t)}-1$ rather than time, we add another equation $\dot{z}=-\left(1+z\right)\frac{\dot{a}}{a}$.
We shall find solutions to this system of equations for given initial conditions. We adjust the clocks in a way that $a(0)=1$, and choose the initial conditions as below\cite{ribas2005fermions}
\begin{equation}\label{initial1}
 \psi_ 1(0) =0.1i, \hspace{5mm}  \psi_ 2(0) = 1, \hspace{5mm}  \psi_ 3 (0) = 0.3,\hspace{5mm}  \psi_ 4 (0) = i,
\end{equation}
where  $\psi_{1}(t)$, $\psi_{2}(t)$, $\psi_{3}(t)$ and $\psi_{4}(t)$ are components of $\psi$, we must specify the initial conditions in such a way that relevant quantities and equations remain real in all the time.
From Friedman equation (\ref{FRW1}) we have
$\dot{a} (0) =a (0) \sqrt{(\rho_ f (0)+\rho_ m (0))/3}$. We also choose $\lambda =0.1$ and $ \rho_ m (0) =2\rho_ f (0)$.

\begin{figure}[h]
\centering
\includegraphics[scale=0.40]{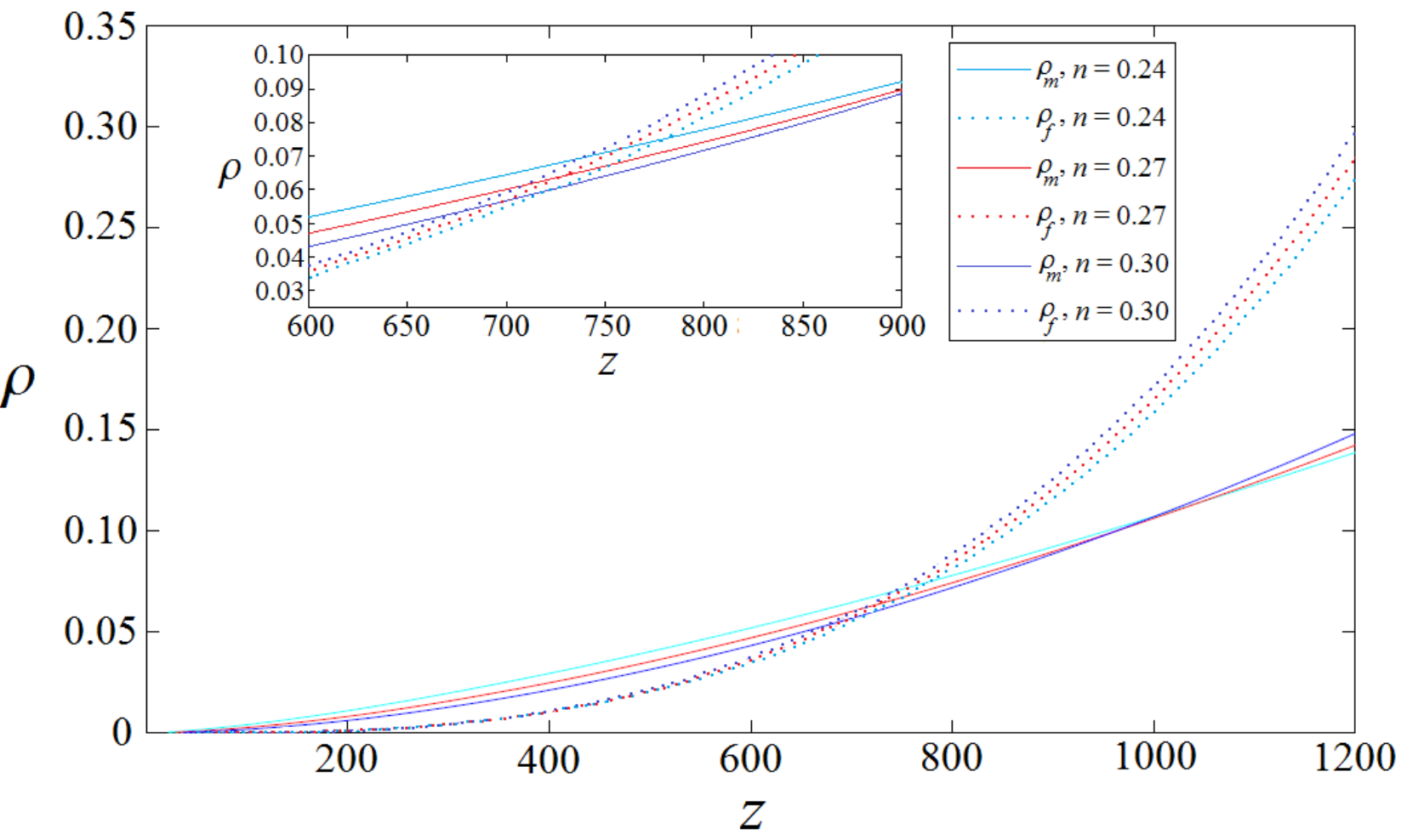}
\caption{The energy densities $\rho_{f}$  (fermionic tachyon field) and $\rho_{m}$  (matter field) are depicted versus red-shift for different values of the exponent n. The initial conditions are indicated in (\ref{initial1}), the other parameters are chosen as m=0.01. At first, for all values of $n$, the matter field is dominated and as time passes(lower red-shifts) the tachyonic field dominates. The figure also shows that for smaller $n$ the matter field decays faster.}
\label{dark density1}
\end{figure}

\begin{figure}[h]
\centering
\includegraphics[scale=0.40]{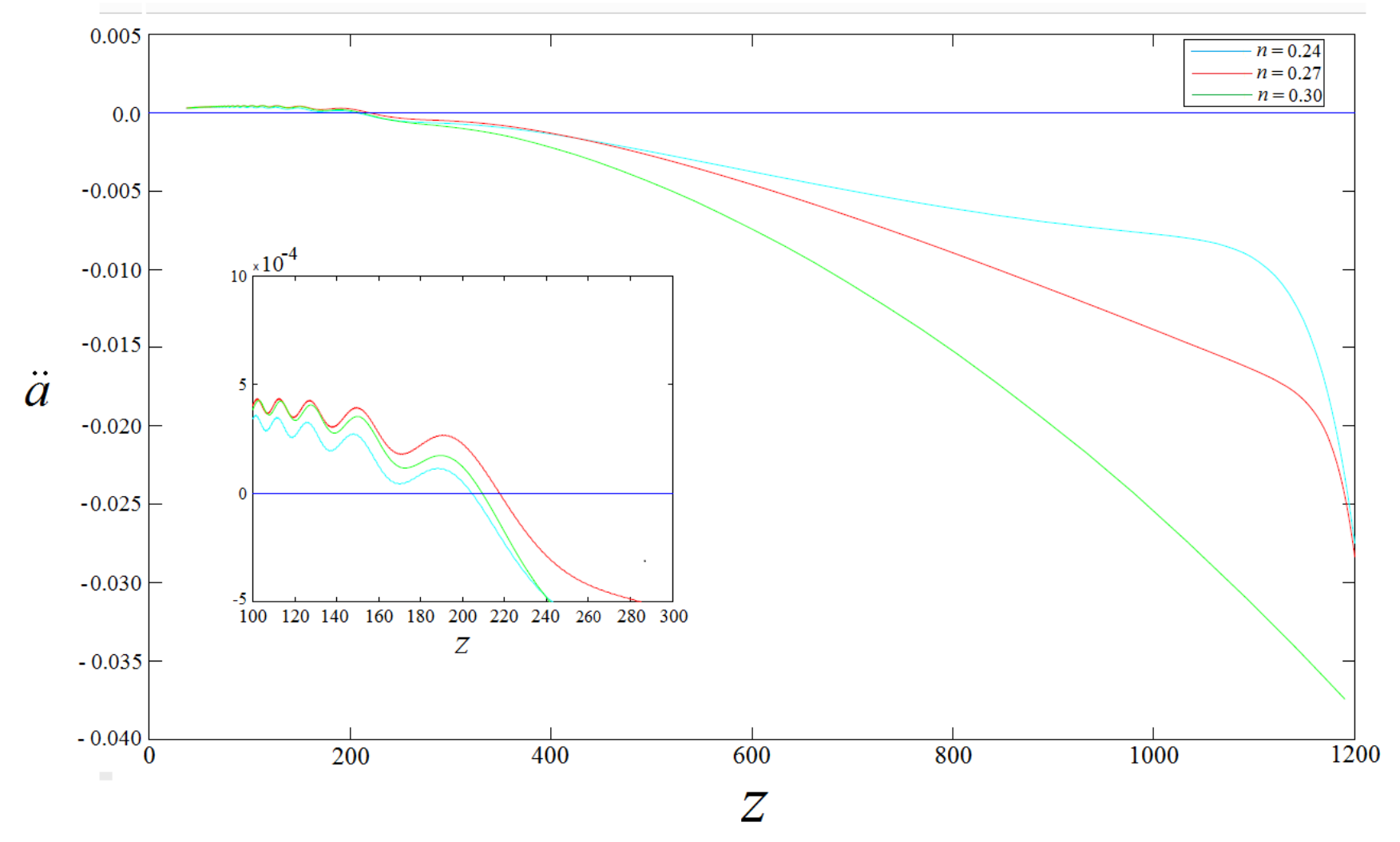}
\caption{The acceleration field $\ddot{a}$ versus red-shift z is demonstrated. The initial conditions and free parameters are the same as in the previous figure. As $n$ becomes smaller, the universe goes into the acceleration era faster.}
\label{dark acceleration1}
\end{figure}
Figure \ref{dark density1} displays the energy densities of fields. Both densities decline with time (increase with z).  After a while,  the fermionic tachyon field dominates and the universe goes into an accelerated regime. That remains valid even if the initial density of matter has been much more than the initial density of the fermionic tachyon. The mentioned accelerating expansion period can be interpreted as the "dark energy" dominated era. In figure \ref{dark acceleration1} the acceleration of universe ($\ddot{a}$) is depicted. At first (large z), since the matter is dominated there is a deceleration era. After a while, the density of dark energy i.e. fermionic tachyon field overwhelms the matter density and the universe goes into an acceleration period. For large values of time(low red-shift), the acceleration declines towards a constant in response to the behavior of the self-interaction potential. It seems that this potential takes the role of the cosmological constant. The exponent $n$ affects the form of the potential and the acceleration. The smaller the value of the exponent, the larger the overall acceleration is achieved. The reduction of $n$ also makes the dark-energy era begin sooner. However, though the model strongly depends on the self-interaction potential, the mass parameter of the tachyonic field has no significant effect on our final results.
 The equation of state for massless tachyons is barotropic; i.e. $p_f=(2n-1)\rho_f$ which gives $\omega_f=2n-1$. So, from (\ref{fermion cons}) in such a situation  $\rho_f\propto 1/a^{6n}$,
and also from \ref{matter cans}  we arrive at $\rho_m\propto 1/a^3$; in comparison to $\rho_f$, it decreases faster with time.

\subsection{The effect of irreversible processes}
  In this section, besides fermionic tachyon and ordinary matter, we take into account the irreversible process of energy transfer between the matter, the gravitational field, and the particle production ~\cite{kremer2002irreversible,kremer2003viscous,kremer2003cosmological,kremer2004acceleration,kremer}.
This process causes a non-equilibrium pressure denoting by $\varpi$.
 Hence the energy-momentum tensor takes the form,
\begin{align}
(T^{\mu}_\nu)=diag(\rho, -p-\varpi, -p-\varpi, -p-\varpi),
\label{energy momentum}
\end{align}
the second Friedmann Equation is modified as
\begin{align}
\frac{\ddot{a}}{a} =-\frac{1}{2}H^2-\frac{1}{2}(p+\varpi).
\label{FRW2m}
\end{align}
The continuity equations are also modified  as
\begin{align}
\dot{\rho}+3H(\rho + p+\varpi) = 0,
\label{consm}
\end{align}
\begin{align}
\dot\rho_m+3H(\rho _m + p_m) = -3H\varpi.
\label{matter cansm}
\end{align}
In the latter equation, the right-hand side is the rate of matter-energy production. The extended (causal or second-order) thermodynamic theory implies that the non-equilibrium pressure, $\varpi$, in a linearized theory satisfies the evolution equation
\begin{align}
\tau\dot{\varpi}+\varpi = -3\eta H,
\label{evol equ}
\end{align}
where $\tau$ denotes a characteristic time and $\eta$ is the bulk viscosity coefficient.
We consider a barotropic equation of state for the matter field, i.e.  $p_{m}=w_{m}\rho_{m}$ with $0\leq w_{m}\leq1$.  We also assume that the coefficient of bulk viscosity $\eta$ and the characteristic time $\tau$ are related to the energy density $\rho$ through $\eta=\alpha \rho$ and $\tau=\eta / \rho$, where $\alpha$ is a constant ~\cite{belinskii1979investigation,di2000qualitative,kremer2003viscous,kremer2003cosmological,kremer2002irreversible}. We choose the initial condition for $\varpi$ as $\varpi(0)=0$.

\begin{figure}[h]
\centering
\includegraphics[scale=0.50]{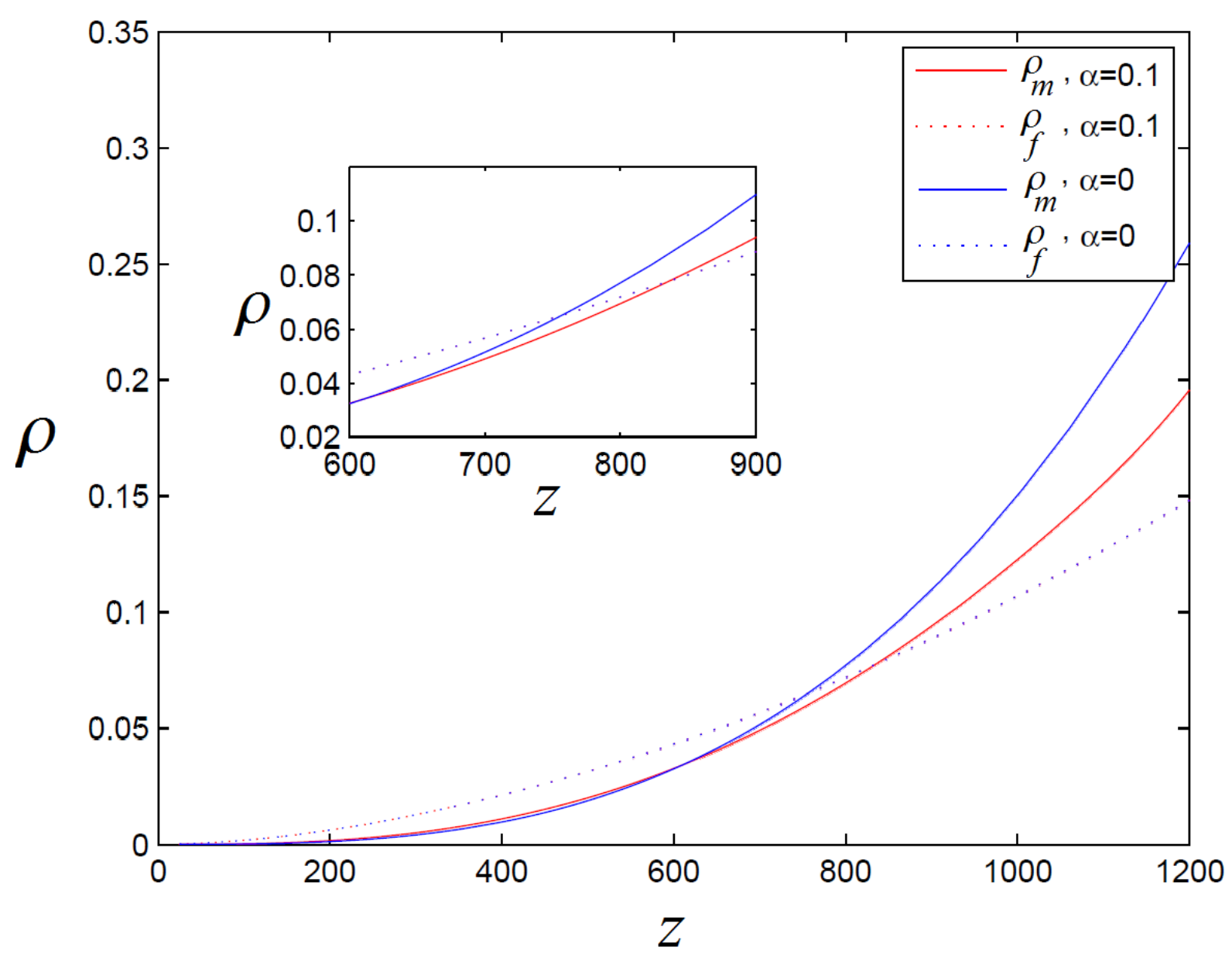}
\caption{ The energy densities $\rho_{f}$  (fermionic tachyon field) and $\rho_{m}$  (matter field) are depicted versus red-shift for two different values of $\alpha$. We set n=0.3, m=0.01 and $\lambda =0.1$. The energy density of the tachyonic field changes pretty the same whether we consider $\alpha$ or not. However, the matter field decays faster when $\alpha$ has a value. Thus considering non-equilibrium pressure only causes a delay in the start of the acceleration era.}
\label{dark density alpha}
\end{figure}

 To obtain the solution, we must replace the modified equations and also add the evolution equation of non-equilibrium pressure (\ref{evol equ}). Figure(\ref{dark density alpha}) shows the effect of the irreversible process on the evolution of density.

\begin{figure}[h]
\centering
\includegraphics[scale=0.50]{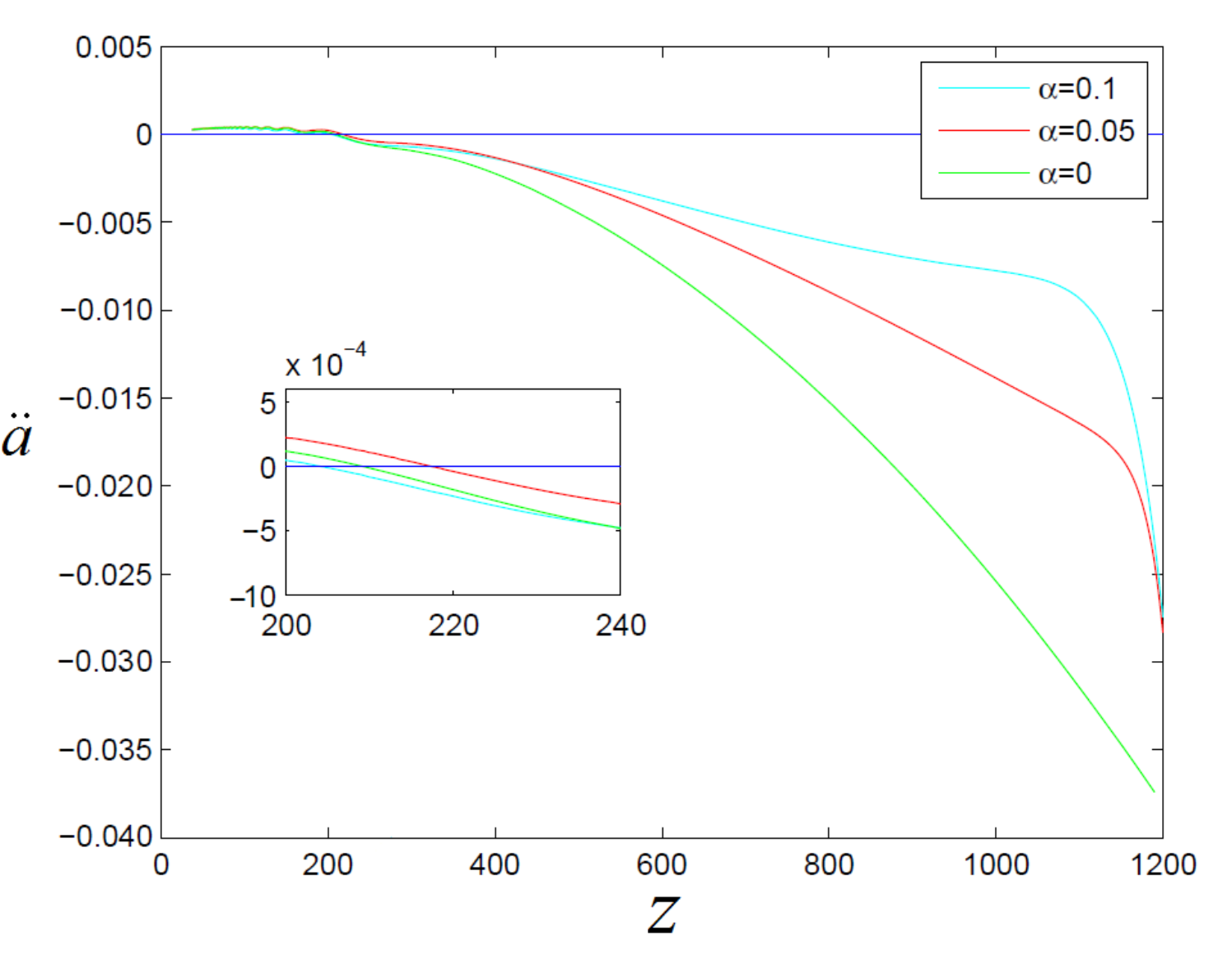}
\caption{The acceleration field $\ddot{a}$  is demonstrated vresus red-shift z for different values of $\alpha$. We set n=0.3, m=0.01 and $\lambda =0.1$. For smaller values of $\alpha$, the acceleration era begins faster. }
\label{dark acceleration alpha}
\end{figure}

It seems that the irreversible processes, deputized by the parameter $\alpha$, slow down the decay of matter.
Changes in the parameter $\alpha$ have a tiny effect on the tachyon density; however, it makes the tachyon-dominated era begin later; see Figs. \ref{dark density alpha} and \ref{dark acceleration alpha}. The presence of viscosity is necessary, as it accounts for the thermodynamic dissipative effects in the expanding universe.  However, it seems that even if we ignore the irreversible processes the main feature of this model would remain almost intact.
\section{ Summary and Conclusion}\label{conclusion}
\begin{figure}[h]
\centering
\includegraphics[scale=0.50]{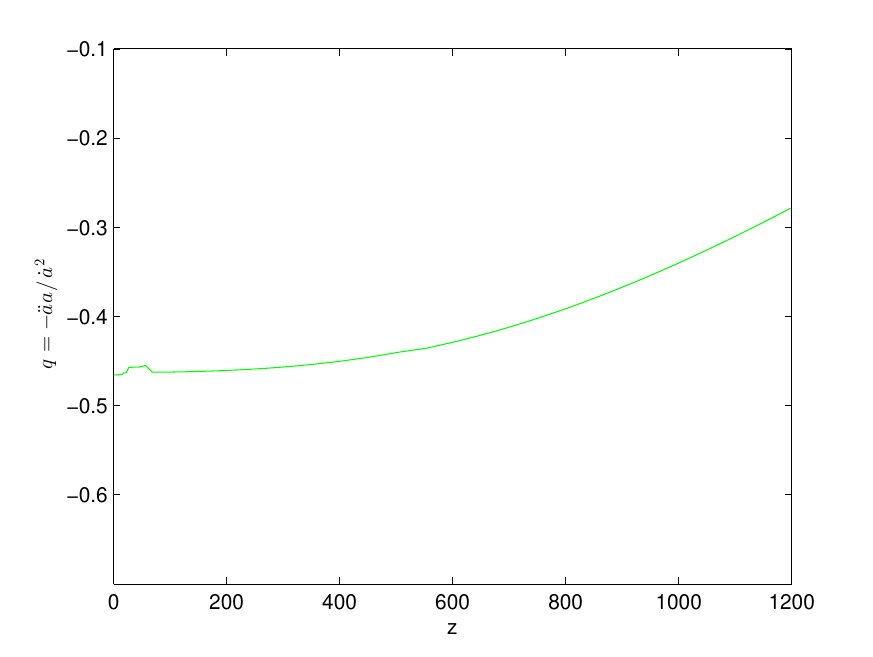}
\caption{The acceleration parameter( we choose $\alpha=0$ $\lambda=0.65$ and  n=0.000009 other parameters are the same as before)}
\label{acceleration parameter}
\end{figure}

With this work, we propose a model in which the fermionic tachyon field can account for the universe's accelerated expansion. It is reasonable to assume that there is a self-interaction potential produced by four-fermion interactions as suggested by effective theories in particle physics. When the mass of the fermionic tachyon is not too large  (we choose m=0.01 in natural units), this potential has a crucial role in producing accelerated expansion. When the potential term is dominated as we expect the mass does not play an important role in our analysis, even if we set $m=0$, there is no significant effect in our main results. We investigate the effect of varying the free parameters of this potential on the behavior of the energy density and the expansion parameter. Our numerical analysis indicates that by reducing the power of the potential "n" and increasing the free parameter "$\lambda$" we get more acceleration, appropriate choosing of these parameters can give enough acceleration in agreement with observation as depicted in  Fig(\ref{acceleration parameter}), at the large time its behavior is similar to cosmological constant means the acceleration approaches a constant value. To be more precise we define dimensionless acceleration parameter q as $q=-\ddot{a}a/\dot{a}^2$ as usual.

Although these results depend on the particular form of our potential, i.e. choosing another form for the potential might change the results. The crucial point of this work is that we have quantitatively shown that it is possible to get accelerated expansion from a fermionic tachyon field.

\section{Acknowledgement}
S.K-M would like to thank IPM astronomy school for hospitality and providing facilities during this work.



\end{document}